\documentclass[useAMS,usenatbib]{mn2e}
\usepackage{graphicx}

\title[Tidal evolution in transiting planetary systems]{Empirical evidence for tidal evolution in transiting planetary systems}

\begin{document}
\bibliographystyle{mn2e}

\author[F. Pont]{Fr\'ed\'eric Pont\\
University of Exeter,
Stocker Road,
Exeter,
United Kingdom.
EX4 4QL}



\maketitle

\label{firstpage}

\begin{abstract}

Most transiting planets orbit very close to their parent star, causing strong tidal forces between the two bodies. Tidal interaction can modify the dynamics of the system through orbital alignment, circularisation,  synchronisation, and orbital decay by exchange of angular moment. Evidence for tidal circularisation in close-in giant planet is well-known. Here we review the evidence for excess rotation of the parent stars due to the pull of tidal forces towards spin-orbit synchronisation. We find 
suggestive empirical evidence for such a process in the present sample of transiting planetary systems. The corresponding angular momentum exchange would imply that some planets have spiralled towards their star by substantial amounts since the dissipation of the protoplanetary disc. We suggest that this could quantitatively account for the observed mass-period relation of close-in gas giants. We discuss how this scenario can be further tested and point out some consequences for theoretical studies of tidal interactions and for the detection and confirmation of transiting planets from radial-velocity and photometric surveys.

\end{abstract}



\section{Introduction}

Tidal interaction in a binary system leads to dissipation and echanges of momentum  that tend, over time, to align the rotation axes of the two components, synchronise their rotation and orbital periods, and circularize the orbit. Evidence of these three effects are clearly observed in close binaries \citep[for a recent review see ][]{maz08}.  \citet{zah77} and \citet{hut81} have estimated the timescales of these processes for pairs of stars. Although the general features of tidal evolution in binary systems are well established, many of the details are still sketchy. Even the main process through which tides affect the orbital elements is not firmly determined, and the key intrinsic parameter $Q$ (tide quality factor)  is expected to vary by orders of magnitude for different objects and different orbital configurations. As a result, both the parametrisation and the absolute scale of the alignement, synchronisation and circularisation timescale are still very uncertain.

In the context of tidal interaction,  systems with a close-in gas giant planet can simply be considered as binary systems with an extremely large mass ratio. Theoretical timescales for tidal evolution from \citet{zah77} and \citet{hut81} are generally used to justify the assumption of a circular orbit when studying close-in transiting planets, or to identify anomalous values of eccentricity or stellar rotation in systems with a close-in planet.    A key difference with stellar binary systems is that, since the mass ratio is so extreme, the synchronisation of the stellar rotation will require a very large inward spiralling of the planetary orbit, through exchange of angular momentum.

Tidal timescales have been calculated specifically for planetary systems in several studies \citep[e.g.][]{ras96,mar02,dob04}, but these papers also stress that timescales in no way represent the whole story, and detailed integration of the dynamical evolution of these systems is necessary. \citet{jac08} have recently carried out such integrations for several known transiting planets, and shown that the tidal evolution varies case by case, and could result in dramatic changes of the orbital parameters over time. \citet{hut80} have established the limit below which an asymptotic equilibrium state towards which the system can evolve no longer exists, and \citet{lev09} show that most known transiting planets lie below this limit. Even systems which satisfy the equilibrium condition may not be stable, because the parent star is not in a steady state. Its rotation is slowed down by magnetic drag, which can counteract the tidal forcing and evacuates angular momentum away from the system. As discussed by \citet{dob04} for instance, if the magnetic drag is large enough, synchronisation is not possible.

 In the absence of an aymptotic equilibrium, the cirularisation and synchronisation timescales lose their conventional meaning, and full numerical simulations are required to determine the evolution of the system. These integrations, though, require a good knowledge of the tidal quality factor $Q$ both for the planet and the star. \citet{mat08} recently reviewed the knowledge of the tidal quality factor $Q$ for close star-planet systems, and concluded that the observed circular orbits would indicate values of $Q$ different by several orders of magnitudes from the values adopted for Solar-System giant planets. Theoretical tidal dissipation models indicate that the $Q$ factor could even vary by orders of magnitudes depending on the frequency of forcing (G. Lesur, priv. comm.). 

For these reasons, we choose in this study to consider the present data on transiting planets for signs of tidal evolution without a priori assumptions from theory, beyond the simple fact that the strength of tidal effects scales with the mass ratio of the system and a steep function of the orbital distance. Besides the well-known evidence for circularisation of planetary orbits, we look for signs of evolution of the stellar rotation towards synchronisation. Spin-orbit alignement is another tidal procees accessible to observations in transiting systems, but the present data is not sufficient for statistical analysis \citep{win08,heb08}. The fourth observable tidal effect -- tidal locking of the planet's rotation -- is out of reach of observations for extra-solar planets.

Obviously, these processes are only relevant if the orbits are not initially circular, aligned and synchronised. In the Solar System for instance , planetary orbits are nearly circular and aligned, not because of tidal effects, but as a relic of initial conditions after formation. The residual eccentricities are sustained by secular interactions with other planets. Formation scenarios of close-in planets by migration in a disc predict initially aligned orbits, with some possible eccentricity injected by interaction with the protoplanetary disc. Scenarios invoking violent dynamical interaction between planets after the dissipation of the disc predict a wider distribution of eccentricities \citep{for99,cha08,nag08}. The present sample of known planets shows that after their formation planets occupy orbits with a very wide range of eccentricities, circular orbits being rare in the absence of tidal effects.

The present status of empirical evidence for tidal effects in close-in planetary system was summarized by \citet{maz08}. 
Evidence for {\it tidal circulatisation} of planetary orbits is obvious in the period-eccentricity relation of the non-transiting exoplanets: planets with periods of a few days predominantly have circular orbits, while at longer periods planetary orbits cover a wide range of eccentricities. This is coherent with timescale calculations, given the leeway offered by the absence of independent constraint on the $Q$ factor of hot Jupiters \citep[e.g.][]{hal05}.
{\it Synchronisation} of the stellar rotation with the planetary orbit can be excluded for all close-in planetary systems from the observed rotation rate of the host star, with a handful of exceptions located near the high-mass end of the planet range, including $\tau Boo$, and HD 162020 \citep{but02,udr02}. This is also compatible with timescale calculations. The star synchronisation timescale for these two system is of the order of the Hubble time, because of the proximity and high mass of the planets. It is much higher than the Hubble time for other systems.

In Section 2, we re-consider the empirical evidence for tidal effects provided by the rapidly growing sample of transiting extra-solar planets. We find tentative but fairly convincing evidence for more indications of tidal evolution. In Section~\ref{sec3} we outline a possible coherent global picture that underlines the importance of tides to understand the orbital characteristics of close-in exoplanets.

\section{Empirical evidence for star-planet tidal interaction}

\subsection{A model-independent look at the transiting planet data}

As discussed in the Introduction, common simplifications to estimate the relevant timescales, and the separation of tidal processes in the star and in the planet, may not be sufficient to predict the effects of tidal interactions on orbital properties. Since the objective of the present study is not to test any individual theory or tidal factors, but to identify empirical evidence for tidal evolution, we attempt to examine the observations in a manner as free of assumptions as possible. 
We identify the systems showing possible signs of tidal evolution, in a simple ranking in terms of the scales that affect the tidal effects. The two main scale factors in tide calculations are the ratio of masses between the two bodies, and the distance between the two bodies compared to their sizes. We take $M_{pl}/M_*$ for the first factor, and $a/(R_{pl}R_*)^{1/2}$ for the second, since the relevant radius is that of the star or planet depending on which effect is considered and how the bodies react to tidal forces \citep{jac08}. We choose this single factor for convenience. Since $R_{pl}$ and $R_*$ vary less than $a$ for the present sample of transiting planets, and $R_{pl}$ and $R_*$ are correlated (bigger stars tend to harbour bigger close-in planets), using $R_{pl}$ or $R_*$ instead produces very similar results. 

The dynamical effects of tidal evolution can all be mimicked by other processes. Circular orbits can result from the planetary formation process, and parent stars can rotate more rapidly because they are younger, independently of tidal spin-up. Indeed, such object-specific explanations have been proposed in many cases for transiting planets. For instance, \citet{al08a} mention young age as a possible explanation for the rapid rotation of COROT-Exo-2, and \citet{rib07} invoke the possibility of an unseen second planet to explain the eccentric orbit of GJ436b. 

 Therefore, the presence of tidal effects has to be examined statistically on ensembles of planets rather than on individual objects. Past the 40-object mark, the present sample of transiting planets has become large enough to start doing this.
 
Table~\ref{table1} gathers the data relevant to the study of tidal evolution for 40 transiting extrasolar planets, from a survey of the literature: the mass and radius of the star and planet, the orbital period, and stellar rotation velocity (from the Doppler broadening of spectral lines\footnote{divided by the sine of the orbital angle, known precisely for transiting planets}) or rotation period (from the periodicity of spot-related variability). The last two quantities are connected through the relation $V_{rot}\cdot P = 2 \pi R$. Rotation velocities derived from spectral lines with an instrumental width larger than the rotational broadening are indicated in parenthesis, since they are more sensitive to systematic uncertainties.

\begin{table*}
\centering
\begin{tabular}{ l l l l l l l  l r r r} \hline
NAME & Mplanet & Rplanet & Porb & a & e & Mstar & Rstar & T$_{eff}$ &Vrot &Prot\\
  & [M$_J$] & [R$_J$] & [days]& [AU] & & [M$_\odot$] & [R$_\odot$] & [K] & [km/s]  & [days]\\ \hline

HD17156 & 3.13  & 1.01 & $\!\!\!$21.22 & 0.15 & 0.67 &1.20  & 1.47& 6079   & (2.6) & \\
HD147506 & 8.04  & 0.98 & 5.63 & 0.067 & 0.50 & 1.32  &1.48   & 6290 & 19.8 &\\
HD149026 & 0.36  & 0.71 & 2.88 & 0.043 & 0 &1.30  & 1.45  & 6147  & 6.0& \\
HD189733 & 1.15  & 1.154 & 2.22 & 0.031 & 0 & 0.82  & 0.755 & 5050 &  & 12.8\\
HD209458 & 0.657  & 1.320 & 3.52 & 0.047 & 0 & 1.10  & 1.12   & 6117& & \\
GJ436 & 0.071  & 0.374 & 2.64 & 0.028 & 0.15 &0.44  & 0.46   & 3200 &     & 48 \\
TrES-1 & 0.76  & 1.081 & 3.03 & 0.039 & 0 & 0.89  & 0.811  & 5250&  &\\
TrES-2 & 1.198  & 1.220 & 2.47 & 0.037 &  & 0.98 & 1.00   & 5850 & (2.0) &\\
TrES-3 & 1.92  & 1.295 & 1.31 & 0.023 &  & 0.90  & 0.80   & 5650 &  &\\
TrES-4 & 0.84  & 1.674 & 3.55 & 0.049 &  & 1.22  & 1.74   & 6200 & 9.5& \\
XO-1 & 0.90  & 1.184 & 3.94 & 0.048 &  & 1.0  & 0.928   & 5750 & (1.1) & \\
XO-2 & 0.57  & 0.973 & 2.62 & 0.037 &  & 0.98  & 0.964   & 5340 & &\\
XO-3 & 13.25  & 1.6 & 3.19 & 0.048 & 0.26 & 1.40  & 1.55 & 6429   & 18.5 & \\
XO-4	&1.72 & 1.34  &		4.13	&	0.055	& &	1.32 & 1.56 & 6397 &	 8.8 & \\
XO-5	&1.15    & 1.15 &         4.19	&0.051	&	&1.00  &	1.11 & 5370& 	1.8 &	\\
HAT-P-1 & 0.53  & 1.203 & 4.47 & 0.055 &  & 1.12  & 1.11 & 5975  & 7.1  & \\
HAT-P-3 & 0.599  & 0.890 & 2.90 & 0.039 &  & 0.94  & 0.824   & 5185 & (0.5) & \\
HAT-P-4 & 0.68  & 1.27 & 3.06 & 0.045 &  & 1.26  & 1.59  & 5860 & (5.5)  & \\
HAT-P-5 & 1.06  & 1.26 & 2.79 & 0.041 &  & 1.16  & 1.17   &5960 & (2.6) & \\
HAT-P-6 & 1.057  & 1.330 & 3.86 & 0.052 &  & 1.29  & 1.46  &6570 & 8.7 & \\
HAT-P-7 & 1.776  & 1.363 & 2.20 & 0.038 &  & 1.47  & 1.84   & 6200 &(3.8) & \\
HAT-P-9	 &0.78  &	1.40	&	3.92		&0.053	& & 1.28 & 6350 &	1.32 	& 12.2  \\
HAT-P-10 & 0.46 &	1.045 &	3.72	&0.044	& &	0.82 &	0.81 & 4980 &	(0.5)& \\
WASP-1 & 0.867  & 1.443 & 2.52 & 0.038 &  & 1.15  & 1.43   & 6110 &(5.8) & \\
WASP-2 & 0.88  & 1.038 & 2.15 & 0.031 &  & 0.79  & 0.81   & 5200 & \\
WASP-3 & 1.76  & 1.31 & 1.85 & 0.032 &  & 1.24  & 1.31   & 6400 &13.4 & \\
WASP-4 & 1.27  & 1.45 & 1.34 & 0.023 &  & 0.90  & 0.953   & 5500& (2.2) &\\
WASP-5 & 1.58  & 1.09 & 1.63 & 0.027 &  & 0.97  & 1.03   & 5700 & (3.4) &\\
WASP-10	&	3.06 &	1.29  &		3.09	& 0.037	& 0.10	& 0.71 	& 0.78 & 4675 &	&	\\
WASP-14	&	7.34 &	1.28 &		2.24 &	0.036	&	& 1.20  &	1.31   &	6475& 	4.9 & \\
COROT-Exo-1 & 3.31  & 1.465 & 1.74 & 0.028 &  & 0.97  & 0.90   & 5950 &(5.2) &\\
COROT-Exo-2 & 1.03  & 1.49 & 1.51 & 0.025 &  & 0.95  & 1.11   &  5625& & 4.6\\
COROT-Exo-4 &	0.72  & 1.19  & 9.20	&	0.090	& &	1.16 	&1.17  & 6190 &	8.9 & \\
OGLE-TR-10 & 0.61  & 1.22 & 3.10 & 0.042 &  & 1.10  & 1.14   & 6075 & &\\
OGLE-TR-56 & 1.29  & 1.30 & 1.21 & 0.022 &  & 1.17  & 1.32   & 6075 & &\\
OGLE-TR-111 & 0.52  & 1.01 & 4.01 & 0.047 &  & 0.81  & 0.83   & 5044 & &\\
OGLE-TR-113 & 1.35  & 1.09 & 1.43 & 0.023 &  & 0.78  & 0.77   & 4804 & &\\
OGLE-TR-132 & 1.14  & 1.18 & 1.69 & 0.030 &  & 1.26  & 1.34   &  6210 & &\\
OGLE-TR-182 & 1.01  & 1.13 & 3.98 & 0.051 &  & 1.14  & 1.14   & 5924 & &\\
OGLE-TR-211 & 1.03  & 1.36 & 3.68 & 0.051 &  & 1.33  & 1.64   & 6325 & &\\ \hline
\end{tabular}
\caption{Parameters relevant to tidal evolution for transiting planets (as of  November 2008). Rotation velocities are measured from spectral line broadening (between parenthesis if smaller than the typical instrumental broadening), and rotation periods from starspots-induced photometric variability.  Uncertainties on masses and radii are generally below the 10\% level. For references see {\it www.exoplanet.eu}.}
\label{table1}
\end{table*}


\label{census}

\subsection{Rotation velocity of planet-host stars}

\label{rot}
Field stars have a distribution of rotation periods that shows a strong dependence with spectral type. Late-type stars (M4 through to G) have long rotation periods (dozens of days) and low rotation velocities (a few km$\, {\rm s}^{-1}$). Early-type stars (F-A) rotate much more rapidly (periods of a few days and below, velocities of several dozen km$\, {\rm s}^{-1}$). The transition is understood to be due to the presence of a convective envelope in late-type stars, that slows down the stars from an initially rapid spin rate through magnetic breaking, combined with the fact that late-type stars are on average older and have had more time to spin down. The exception to this rule are young, late-type stars, which conserve part of their initial rotation momentum, in inverse relation to their age, as well observed in star clusters. Among solar-neighbourhood stars, such young, rapidly-rotating late-type stars are rare \citep[less than 5\%,][]{nor04}.

\begin{figure*}
\resizebox{16cm}{!}{\includegraphics{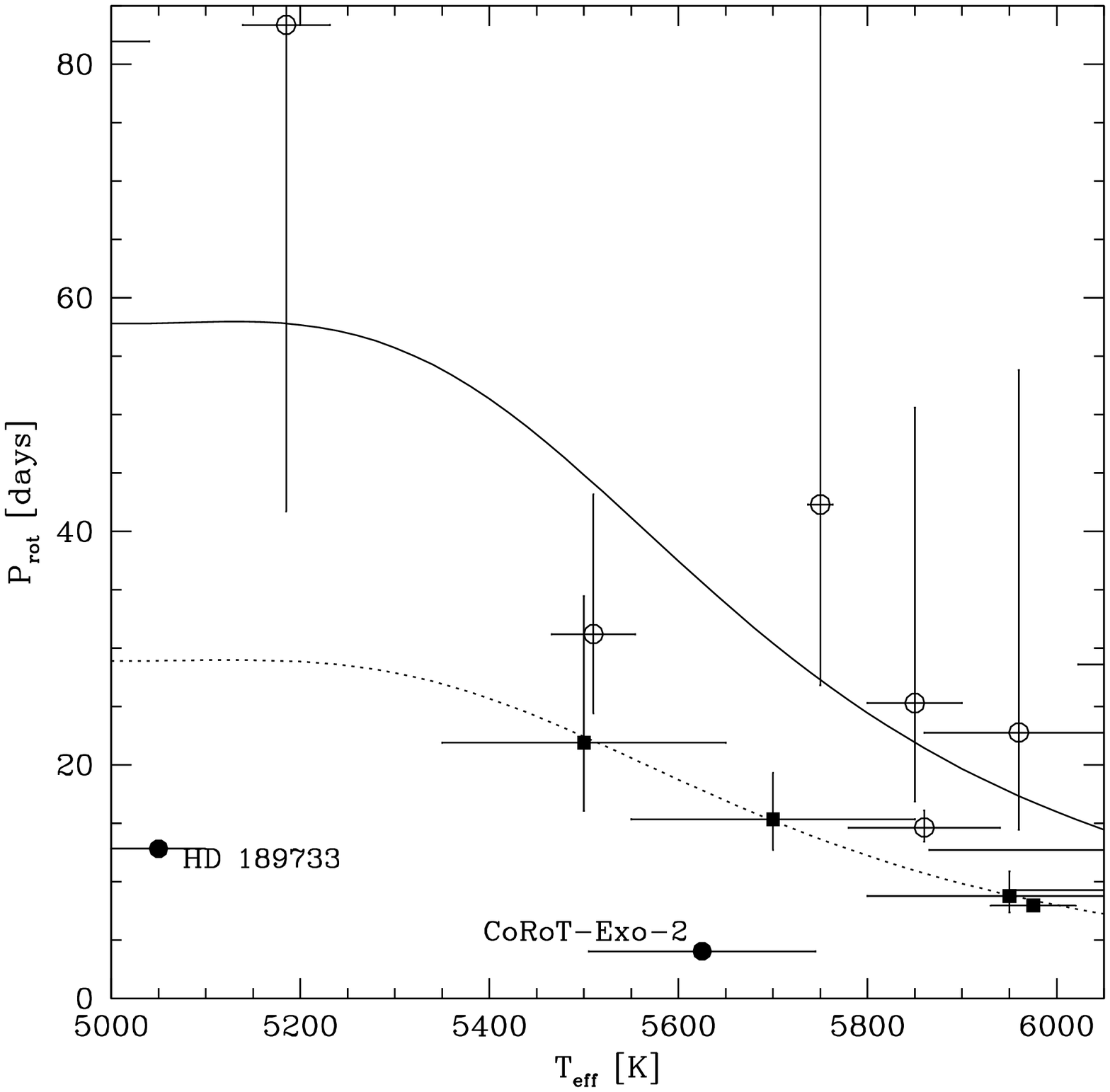}\includegraphics{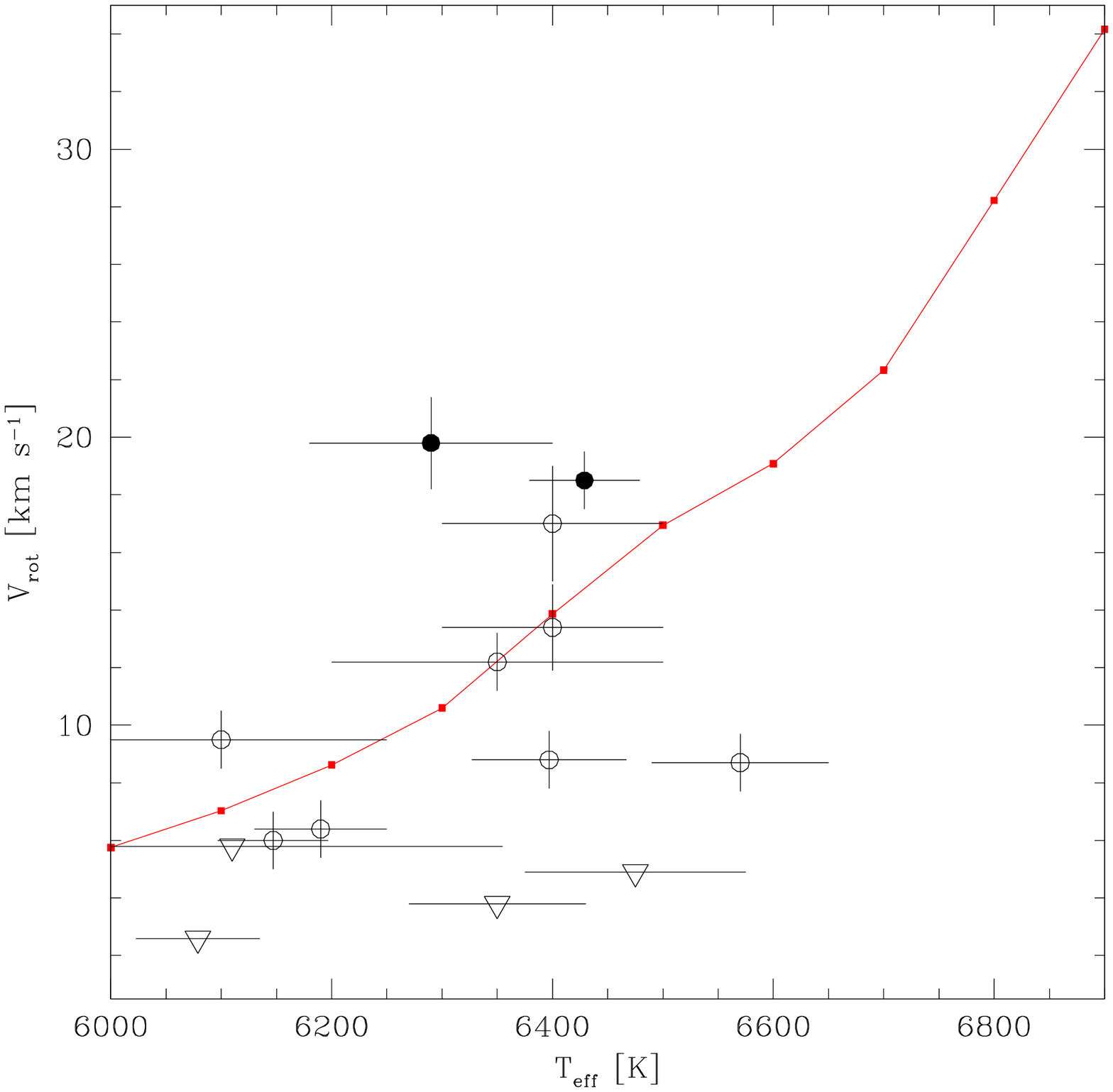}}
\caption{{\bf Left:} Rotation period vs temperature for late-type transiting planet host stars, with 4.6 Gyr isochrone from \citet{str98} (solid line), and half this rotation period (dotted line). {\bf Right} Rotation velocity of early-type transiting planet host stars, with median relation for the Solar-Neighbourhood sample of \citet{nor04}. Triangles indicate rotation velocities below the instrumental resolution. The four objects with large excess rotation are indicated with filled circles (HD 189733 and Corot-Exo-2 among late-type stars and HD 147206 and XO-3 among early-type stars). Four objects with possible excess rotation are marked with squares. }
\label{figvrot}
\end{figure*}

The rotation velocity of stars is best measured in different ways for late-type or early-type stars. In F-dwarfs, the rotation is rapid enough that the spectral lines are broadened beyond the instrumental broadening of  high-resolution spectrographs, so that the projected rotation velocity $V_{rot} \sin i$ can be measured from the width of the spectral cross-correlation function. 
For G and K dwarfs, the rotational broadening is usually below the instrumental broadening, and the rotation velocity can be recovered only with limited accuracy -- typically $\sim \!\!1$ km$\,{\rm s}^{-1}$ uncertainty on a value of a few km$\,{\rm s}^{-1}$. But the stellar rotation period can be measured with sufficiently precise photometric monitoring, by measuring the quasi-periodic brightness modulations due to the motion of starspots across the stellar surface. This has been done for instance in the cases of HD 189733 \citep{win06,cro07} and GJ 436 \citep{dem07}. In transiting planetary systems, the $\sin i$ and $R_*$ are known with sufficient accuracy, so that $P_{rot}$ can be translated into $V_{rot}$.

Figure~\ref{figvrot} shows the rotation velocity (for F stars) or period (for K-G stars) information for transiting planet host stars. The reference relation for F stars is taken from the median of the distribution of rotation velocities in \citet{nor04} in 100 K temperature bins. The mean relation for K-G stars is taken from the 4.6 Gyr  isochrone of rotation periods from \citet{str98}.

From this figure, four systems are identified as having excessive rotation for their spectral type, HD 189733, Corot-Exo-2b, HD 147506, XO-3. Another four systems are above the mean relation, HAT-P-1, WASP-4, WASP-5, \mbox{CoRoT-Exo-1}. Late-type objects that have rotation periods longer than the 4.5 Gyr isochrones are flagged as showing normal rotation, while early type stars ($T_{\rm eff}>6000 $K) without significant excess rotation are flagged as ``undetermined'', since their intrinsically high rotation could hide a significant componant from tidal spin-up.

\subsection{Empirical evidence of tidal excess rotation}

\begin{figure*}
\resizebox{16cm}{!}{\includegraphics{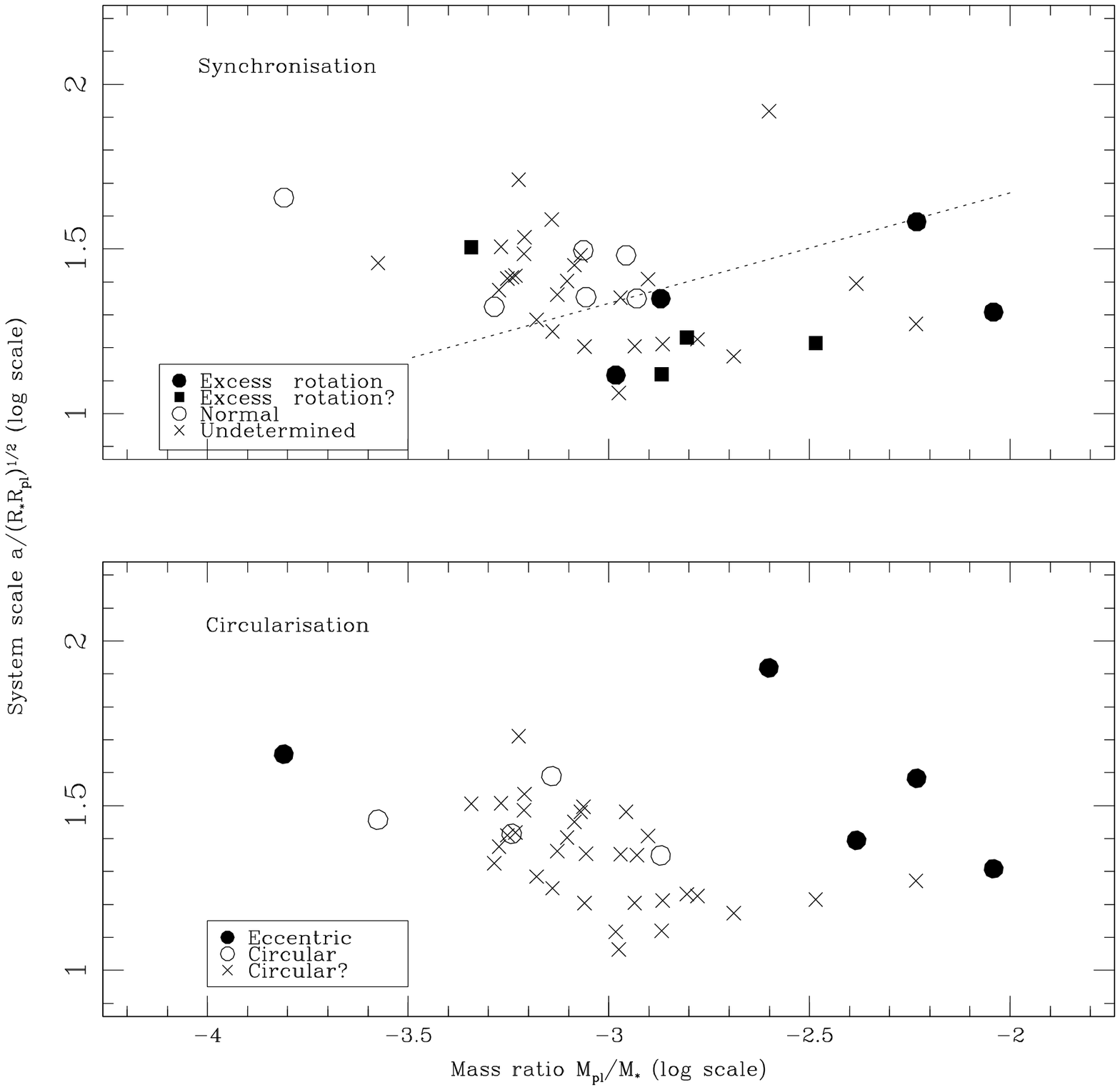}}
\caption{Mass ratio vs. system "scale" $a/(R_*R_{pl})^{1/2}$ for transiting planets. {\bf Top:} Filled symbols mark host stars with definite excess rotation (circle) and possible excess rotation (square), open circles systems with normal rotation, and crosses early-type stars without definite excess rotation; the line indicates the slope of constant scale for tidal effects. {\bf Bottom:} filled circles mark eccentric orbits ($e>0.1$), open circles indicate circular orbits ($e<0.05$). Crosses indicate cases without precise measurements:  apparently negligible but poorly constrained eccentricity.  }
\label{tides}
\end{figure*}

The upper panel of Figure~\ref{tides} shows the position of the transiting systems in terms of the scaling of tidal effects (from Par.~\ref{census}). Objects with characteristics relevant to tidal evolution towards synchronous rotation of the star are indicated with special symbols (from Par.~\ref{rot}). The scale of tidal effects by the planet on the star increases towards the lower left of the plot. The dashed line shows the slope of constant synchronisation timescale  according to \citet{zah77},  $\tau_{\rm synch}\sim (a/R)^6 (M_{pl}/M_{*})^{-2}$. 

The plot shows a highly non-random relation between possible excess rotation and ranking in terms of synchronization timescale, suggesting a possible causal relationship. Because the tidal synchronisation timescale is much higher than $10^{10}$ years for all these object, tidal spin-up has  been dismissed as a possible cause of the high rotation of, for instance, HD 189733 \citep{bou05} and Corot-Exo-2 \citep{al08a}, and young ages preferred as a possible cause of excess rotation. However, the ensemble position of host stars with excessive rotation in Figure~\ref{tides} makes it very likely that tidal effects have played a dominant part in a majority of these systems to explain the faster rotation than expected. 

With hindsight, this is less surprising: the synchronisation timescale estimates the time needed to reach the asymptotic equilibrium stage, with the star turning as rapidly as the planetary orbit. But tidal forces can still affect the rotation of the star in a detectable way by compensating part of the magnetic drag and slowing down the secular decrease of the stellar spin. 

\subsection{Non-transiting planets and excess rotation}

\begin{figure}
\resizebox{8cm}{!}{\includegraphics{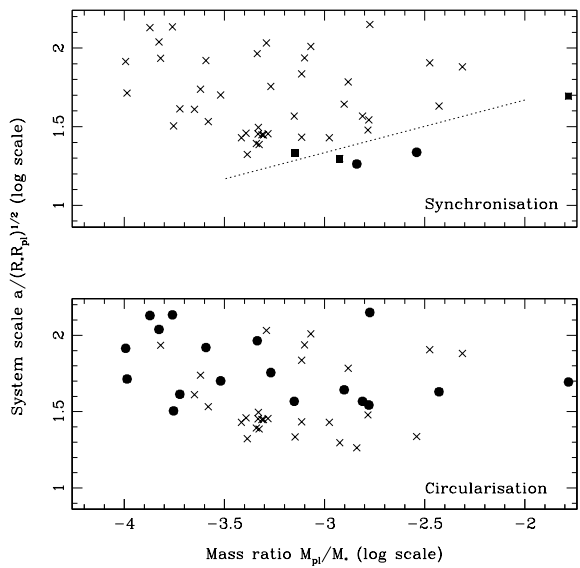}}
\caption{Mass ratio vs. system scale for non-transiting  planets. Axes, symbols and line as in Fig.~\ref{tides} }
\label{Doppler}
\end{figure}

The tidal parameters are less well determined for non-transiting planets, because of the $\sin i$ indeterminacy on the stellar rotation velocity, and the less precise measurements of the stellar and planetary mass and radius. Nevertheless, close-in non-transiting planets offer a chance of confirmation of the presence of tidal excess rotation.  We can produce a first-order ``tidal mass-distance plot'' for non-transiting planets. We use $R=1.2$ R$_{\rm J}$ for planets heavier than Saturn, and an interpolation of the mass-radius relation defined by Solar-System planets in a log-log plane for lighter planets. The results are plotted on Fig.~\ref{Doppler}. Five objects have tide strength values in the range of the transiting systems with possible excess rotation: $\tau Bootis$, HD 73256, HD 86081, HD 179949 and HD 162020. The rotation rate of the five host stars clearly support the hypothesis of tidal evolution. Two of them have spin rates compatible with synchronous rotation, the heavy-planet hosts $\tau Boo$ and HD 73256. Two others are also probably synchronised, although the low $\sin i$ makes it difficult to tell  \citep[Udry et al. 2002 for HD 162020;][for HD 179949]{saa98} , and the last has a slightly too high rotation velocity for its spectral type \citep[][ for HD 86081, $V {\sin i}=4.2 \pm 0.5$ km/s]{joh06}.  
Thus the position of objects with possible excess rotation and synchronisation appears non-random as well for non-transiting planets, and coherent with the relevant panel of Fig.~\ref{tides} for a single transition locus for significant spin-up.

Note that the detection of planets by the radial velocity method is very strongly biased towards slow-rotating parent stars. Doppler survey usually weed out fast-rotating targets, identified either by the broadening of the spectral lines about the instrumental resolution, or by activity-induced variability. Comparing the corresponding panels of Figures~\ref{tides} and~\ref{Doppler} illustrates the radial velocity bias compared to the transit method (see Section~\ref{stats}).

\subsection{Evolution of the orbital distance}

In order to conserve the total angular momentum of the system, 

\[
{\cal L}_{tot} = {\cal L}_{*} + {\cal L}_{pl} = \frac{I_*}{M_* R_*^2} M_* R_* V_{rot} +  M_{pl} a V_{orb} \, ,
\]  
tidal excess rotation of the star must lead to a shortening of the orbital period (we neglect the contributions from planetary rotation and orbital circularisation). Therefore, under the assumption that, for the objects identified above, the stellar excess rotation is due to tidal interaction, one can recover the initial orbital distance, by putting the angular momentum of the excess rotation back in the planetary orbit. This will be an {\it underestimate}, since part of the angular momentum imparted to the star will be dissipated by the extra magnetic breaking caused by the faster spin.  

Figure~\ref{mp} shows the resulting minimal initial orbital distance obtained in this way for the two objects with precisely measured rotation velocity, HD 189733 and Corot-Exo-2, assuming the moment of inertia of the Sun ($I/MR^2=0.06$) for host stars.  Although only two  ``hot Jupiters'' can be analysed in this way, the result is extremely suggestive: computing the initial distance places these two objects back at the typical distance of less massive hot Jupiters, therefore exactly compensating the observed mass-period relation shown by\citet{maz05}. This suggests that the mass-period relation may be entirely due to secular tidal spin-up and orbital decay for the heaviest hot Jupiters, after an initial mass-indifferent pile-up near $P=3$ days. 

This scenario has another important consequence: companions ending up near $P=3$ days with masses near 2 $M_{Jup}$ and higher would have a strong enough coupling with the star to spiral inwards all the way to the Roche limit within the lifetime of the star, a scenario probably resulting in their destruction. At higher masses ($M_{pl}>\frac{I_*}{M_* R_*^2} M_*R_* V_{\rm synch}/a V_{\rm orb}^{\rm init}$), the heaviest planets will have enough angular momentum to bring the star to synchronisation before spiralling all the way into the star, and reach a  tidally locked state, provided their starting position lies beyond a certain critical distance.

\subsection{Evidence for tidal  circularisation}

Although initially unexpected, the wide range of planetary eccentricities is now well established \citep[e.g.][for a recent discussion]{hal05} from the large sample of planets detected by radial velocity. The effect of tidal circularisation is also clearly seen: orbits shorter than 6 days are usually circular.  

The exceptions -- very short period, eccentric orbits -- are often attributed to possible eccentricity-pumping by an unseen planetary companio . However, it is remarkable that on the lower panel of Figure~\ref{tides} that the transiting planets with markedly eccentric orbits are all situated at the outer edge of the sample in terms of scales relevant to tides.  Figure~\ref{Doppler} shows the same plot for non-transiting planets.  The location of planets with eccentric orbits in this plane is also compatible with a single limit. In the most populated part of the parameter space, this limit it comparable to that defining excess stellar rotation -- suggesting a separation between systems strongly affected by tidal evolution and others.  

Consequently, an extraneous cause for eccentric orbits is not required for any planet, and  in all cases the eccentricity could be a relic of the initial orbit after formation, possibly decreased by incomplete circularisation, without the need to invoke the influence of another planet or star. Remarkably, this is also the case for GJ436b, whose short orbital period is compensated by the small scale of the system. The very low mass of GJ 436b suggests that the tidal quality factor $Q$ may be in a significantly different regime than gas giant planets. Therefore, while the existence of an eccentric orbit at such a small orbital distance (2.6 days) is surprising at face value and has been thought to require an additional planetary companion, when the system is scaled in size, it turns out that the tidal effects are comparable to other cases that have kept their eccentric orbit. Therefore the eccentricity of GJ436 could be a relic of formation.

\subsection{Relation with planet size}

One of the most remarkable feature of the present sample of transiting planets is the large spread in radius of hot Jupiters, and the presence of several planets with sizes much larger than predicted by structure models. Several types of explanations have been proposed to account for these "bloated" hot Jupiters, such as increased opacities \citep{bur07}, transformation of part of the incident stellar flux into mechanical energy \citep{gui06}, and tidal heating, with \citep{win06} or without \citep{jac08} requiring the presence of an additional planet. 

In relation to the present study, tidal effects like orbital circularisation inject internal energy into the planet, and therefore delay its contraction. \citet{jac08} have estimated the amount of tidal energy absorbed by the planet during orbital circularisation, and show that it is large enough to affect the planetary radius. 

\begin{figure}
\resizebox{8cm}{!}{\includegraphics{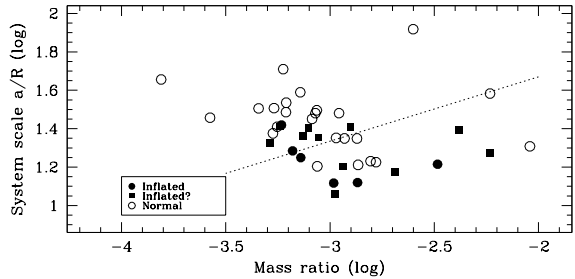}}
\caption{Mass ratio vs. system "scale" $a/(R_*R_{pl})^{1/2}$ for transiting planets.  Closed symbols mark planets with large excess radius (circles, $R_{pl}\!-\!\sigma_R\!  >$1.3  R$_{\rm J}$) and moderate excess radius (squares, 1.3 R$_{\rm J} \!>\! R_{pl}\!-\!\sigma_R \!>$1.2  $_{\rm J}$). }
\label{bloat}
\end{figure}

\begin{figure}
\resizebox{8cm}{!}{\includegraphics{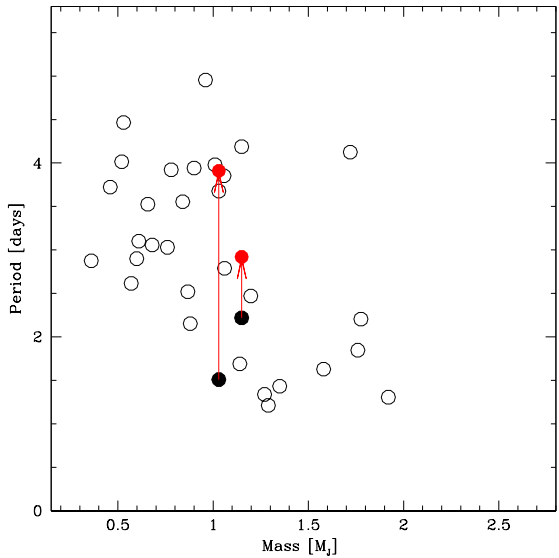}}
\caption{Mass-period relation for transiting hot Jupiters. The black dots are the two system with precisely quantified excess spin of the star, with the arrows and upper dots showing the initial period if the angular moment is restituted to the planetary orbit.}
\label{mp}
\end{figure}

We repeat our analysis of tidal scales, identifying planets with anomalously large radii, by comparison of their observed position in the mass-radius diagrams with theoretical models from \citet{bar08}. The result is shown in Fig.~\ref{bloat}, on the same plane as Fig.~\ref{tides}.

If tidal effects were related to radius excess, some correlation would be expected between the strength of tidal forces and the excess radius. Correlation does not prove causality, however, and the link can be indirect. For instance, \citet{fre07} recently pointed out the strong correlation between excess radius and the incident stellar flux, supporting the scenario of \citet{gui06}. Incident flux, like tidal effects also scales with $a/R$.

Fig.~\ref{tides} shows that, in the present sample, the correlation of excess radius is stronger with $a/R$ than with the strength of tidal effects. The present sample of transiting planet is not sufficient to distinguish between the two types of explanations without further modelling, but it does suggest that tidal effects are not the main factor in determining the size of close-in gas giant planets and that the intensity of the incident flux is a better candidate.

\section{Tidal evolution in the mass-period plane}
\label{sec3}

If the tentative indications provided by the present sample of transiting planets are confirmed, the following overall picture emerges:\\
\ \\
- close-in planets end up the formation phase on orbits with a wide distribution of eccentricity,\\
\ \\
- tidal interaction with the star circularise the planetary orbit, beginning around $log(a/R)=1.5$ for 2-M$_{\rm J}$ planets,  then circularizes it. Early signs of circularisation are visible already at larger distances,\\
\ \\
- for planets lighter than Jupiter, the circularisation is not accompanied by significant orbital decay and effect on the star,\\
\ \\
- for planets with mass comparable  to Jupiter or slightly heavier, circularisation is accompanied by significant orbital decay, leading to a mass-period relation, and possible destruction for $M\sim 1-2 M_{\rm J}$,\\
\ \\
- still heavier planet produce a marked stellar excess rotation, and corresponding orbital decay, during the process of tidal evolution and circularisation. If their initial position exceeds a critical value, they may reach a spin-orbit synchronous state with the star,
 \\ \\
- as a result of this evolution, the orbital characteristics of very close-in planets (``hot'' planets) are profoundly modified by tidal interactions, including the orbital distance and period. Planets lighter than 2 $M_J$ and closer than a few days are "sheperded" along a mass-period trend controlled by tidal angular momentum exchange and possibly tidal inflation of the planet. The period-mass relation of close-in planets cannot be used at face value to constrain formation and migration scenarios.\\
\ \\

\begin{figure}
\resizebox{8cm}{!}{\includegraphics{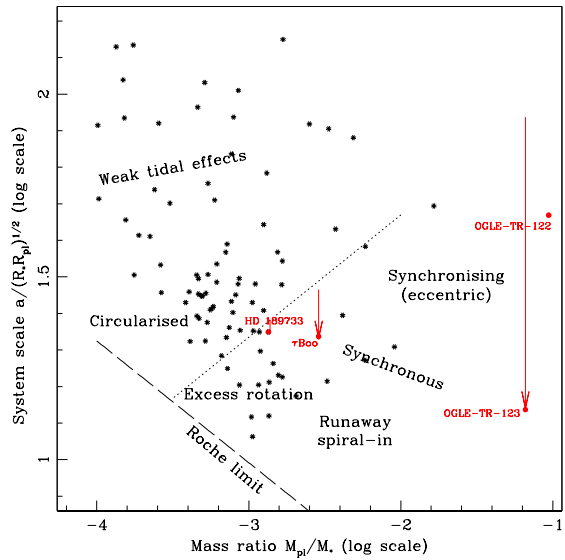}}
\caption{"Tidal" mass-period diagram for gas giant exoplanets, brown dwarfs and small stars: Orbital distance in units of star and planetary radius vs.  mass ratio. The dotted line delienates the possible transition zone to detectable spin-up and orbital decay   (see text). The crosses mark known planets. Four interesting objects are labelled, the arrows show the orbital evolution if the present angular momentum of the star is placed back in the planetary orbit.}
\label{am}
\label{fig5}
\end{figure}

If this interpretation of the data is correct, we can roughly divide the mass-period plane for exoplanet according to their sensitivity to tidal evolution. 
Figure~\ref{fig5} sums up the observable effects of tidal evolution as inferred here from transiting planets. The possible transition zone where orbital circularisation and  orbital decay occur is indicated, as well as the Roche limit (with reasonable assumptions on the mass-radius relation of planets). 

Planets at the weak end of the scale of tidal effects can conserve eccentric and misaligned orbits. Jupiter-mass planets placed near the 3-days limit will substantially affect the rotation of their parent star, moving closer in the process by orbital decay, down to 1-2 day periods. Planets near the 2-3 Jupiter mass range will undergo strong orbital decay, facing potential destruction if they start on a close orbit.  Large planets or brown dwarfs in the 5-15 Jupiter-mass range will spin up their host star substantially if they orbit close enough, all the way to synchronous rotation. Lighter planets will not affect their star detectably. If they are close enough, they will spiral inwards to the Roche limit in a short time. 

As more objects fill the diagram, this global interpretation can be tested and refined. 

In Figure~\ref{fig5}, the two known transiting stars with the lowest masses are also plotted, OGLE-TR-122 \citep{pon05} and OGLE-TR-123 \citep{pon06}. These objects show that the  connection with stellar binaries is coherent, with the first object having an orbit still eccentric but clear evidence of excess rotation of the primary, and the second  a circular and synchronous orbit.

Figure~\ref{fig5} shows, for HD~189733, $\tau$ Boo and OGLE-TR-123, the initial orbital distance assuming that all the angular momentum of the star is put back in the orbit. The initial positions will be placed near the limit of tidal spin-up, which provides additional support to this interpretation. It also suggests one of the reasons for the inaccuracy of tidal timescales calculated from present orbital parameters: the initial orbit could have been very different.  For OGLE-TR-122 for instance, the initial position is above the frame of the plot.

From this figure, several predictions and possible tests of our tentative interpretation of the data emerge: 
\begin{itemize}
\item{the rotation period of host stars with planets in the spin-up zone should be shorter than average. This includes TrES-3, OGLE-TR-56, OGLE-TR-113}
\item{planets in the rapid orbital decay zone will be rare ($M> 2M_J$, $P<2$ d)}
\item{planets at the edge of the circularisation zone may have residual small but detectable eccentricities, namely OGLE-TR-132, XO-1, OGLE-TR-111, OGLE-TR-211, HAT-P-3}
\item{seemingly ``young'' late-type stars will be more abundant among close-in host stars from transit surveys than in the general field}
\item{many heavier hot Jupiters were missed by Doppler surveys around late-type targets because of the rapid rotation of the primary.}
\end{itemize}

\section{Consequences for planet detection and statistics}

\label{stats}

Tidal evolution is related to a strong selection bias in exoplanet detection: rotating stars have broadened spectral lines, and the detection of planetary-amplitude Doppler reflex motion is increasingly difficult. Moreover, late-type fast rotators are usually photometrically active, because of magnetically generated star-spots, which produces asymmetries in the line shapes that can mimic planetary orbits. For this reason, stars with excess rotation are usually dropped from planet searches, producing a strong bias against systems with tidal spin-up. 

Photometric transit searches suffer a smaller bias in this respect. The spectroscopic follow-up is usually carried out even in the presence of stellar rotation. Still, systems showing rotation velocities compatible with tidal synchronisation or photometric effects showing rotation of the primary with a period comparable to that of the transit signal, are often dismissed from transit planet searches as probable eclipsing binaries. 

If the interpretation presented in this paper is correct, we would predict the existence of close-in planets that have substantially affected the rotation of their star, and escaped detection as a result. Specifically, there may be undetected planets at the high end of the mass range (5-15 $M_J$), with spin-orbit synchronisation. These could be detected by including late-type stars in rapid rotation in normal planet searches, since the high amplitude of the radial-velocity orbit will partly compensate the lower precision. The importance of this selection bias may be worth investigating in more quantitative details. 

We also predict planets near the 1-day, 2 $M_J$ mark, that have pulled their host star somehow towards synchronisation. The example of HD 189733 shows that slightly active and rotating stars should not be excluded from Doppler planet searches before checking for the presence of a high-amplitude, short-period orbital signal. This may explain the 2-sigma incompatibility between the abundances of very short-period hot Jupiters from transit surveys and Doppler searches \citep{gau05}.

\begin{figure}
\resizebox{8cm}{!}{\includegraphics{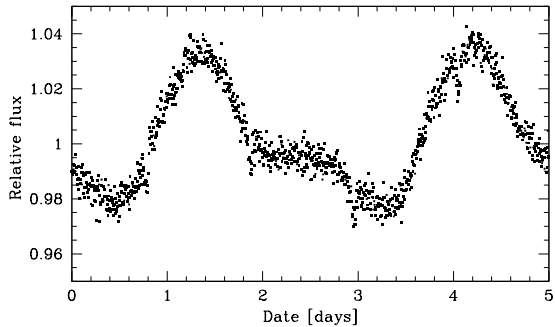}}
\caption{Possible light curve of an imaginary transiting planet with $P\sim 0.9$ days and $M\sim 2 M_J$, having imparted significant angular momentum to its solar-type host star and spiralled inwards, as would be measured with a continuous sampling of 6 minutes and a noise of $3\cdot 10^{-3}$ on individual points (built from the lightcurve of CoRoT-Exo-2).}
\label{fig4}
\end{figure}

Another implication is relevant to photometric transit searches: our interpretation predicts the existence of planets undergoing strong orbital decay, with host stars rotating rapidly. This kind of system will not fit the assumptions of standard photometric transit search and may be missed by detection algorithm and manual candidate selection. They will show rotation-induced variability on short timescales, and a very short-period transit signal, with a long duration in phase. Figure~\ref{fig4} shows the possible lightcurve of such a putative system, obtained by scaling the case of Corot-Exo-2b \citep{al08a}, a system with intermediate orbital decay and stellar excess rotation. 

\section{Summary and conclusions}

We have examined the set of known transiting extrasolar planets as a whole for signs of tidal evolution. We find evidence for well-defined limits in parameter space beyond which the orbits of planets are circularized, and beyond which the host star shows sign of excess spin due to tidal effects, roughly following the dependence expected from simple scale arguments. We find that all presently observed eccentric orbits and fast-rotating host stars for transiting planets are compatible with tidal evolution -- and that alternative explanations in terms of 3-body mechanisms or young stellar ages are not required. The data for non-transiting close-in planets, although less precise, confirm this global picture.

These tentative conclusions would imply that below a certain distance ($P\sim 3-5 $ days), the orbital properties of planets are controlled by tidal effects. Not only the orbital eccentricity, but also the orbital distance and stellar spin, can be very different from the ones set after planetary formation. In order of increasing mass, we would expect close-in planets to be unaffected ($M<<M_J$), circularized ($M\sim M_J$), spiralling in ($M \sim 1-2 M_J$), destroyed ($M \sim 2-3 M_J$), synchronised ($M > \sim 3 M_J$). The observed mass-period relation of hot Jupiters \citep{maz05} would be a footprint of this mass-dependent "sheperding" of close-in planets.

One important practical implication is that tidal spin-up of the host star will produce a strong detection bias against massive, close-in planets, especially in Doppler surveys. 

\section*{Acknowledgements} The author warmly thanks Gilles Chabrier, Isabelle Baraffe, Tsevi Mazeh, Eric Ford, Josh Winn and Genya Takeda for illuminating discussions and comments, and the anonymous referees for very helpful suggestions.

\bibliography{tidal}{}

\begin{thebibliography}{}

\bibitem[\protect\citeauthoryear{{Alonso}, {Auvergne}, {Baglin}, {Ollivier},
  {Moutou}, {Rouan}, {Deeg}, {Aigrain}, {Almenara}, {Barbieri}, {Barge}, {Benz}
  \& {Wuchterl}}{{Alonso} et~al.}{2008}]{al08a}
{Alonso} R.,  {Auvergne} M.,  {Baglin} A.,  {Ollivier} M.,  {Moutou} C.,
  {Rouan} D.,  {Deeg} H.~J.,  {Aigrain} S.,  {Almenara} J.~M.,  {Barbieri} M.,
  {Barge} P.,  {Benz} W.,    {Wuchterl} G.,  2008, \aap, 482, L21

\bibitem[\protect\citeauthoryear{{Baraffe}, {Chabrier} \& {Barman}}{{Baraffe}
  et~al.}{2008}]{bar08}
{Baraffe} I.,  {Chabrier} G.,    {Barman} T.,  2008, \aap, 482, 315

\bibitem[\protect\citeauthoryear{{Bouchy}, {Udry}, {Mayor}, {Moutou}, {Pont},
  {Iribarne}, {da Silva}, {Ilovaisky}, {Queloz}, {Santos}, {S{\'e}gransan} \&
  {Zucker}}{{Bouchy} et~al.}{2005}]{bou05}
{Bouchy} F.,  {Udry} S.,  {Mayor} M.,  {Moutou} C.,  {Pont} F.,  {Iribarne} N.,
   {da Silva} R.,  {Ilovaisky} S.,  {Queloz} D.,  {Santos} N.~C.,
  {S{\'e}gransan} D.,    {Zucker} S.,  2005, \aap, 444, L15

\bibitem[\protect\citeauthoryear{{Burrows}, {Hubeny}, {Budaj} \&
  {Hubbard}}{{Burrows} et~al.}{2007}]{bur07}
{Burrows} A.,  {Hubeny} I.,  {Budaj} J.,    {Hubbard} W.~B.,  2007, \apj, 661,
  502

\bibitem[\protect\citeauthoryear{{Butler}, {Marcy}, {Vogt}, {Tinney}, {Jones},
  {McCarthy}, {Penny}, {Apps} \& {Carter}}{{Butler} et~al.}{2002}]{but02}
{Butler} R.~P.,  {Marcy} G.~W.,  {Vogt} S.~S.,  {Tinney} C.~G.,  {Jones}
  H.~R.~A.,  {McCarthy} C.,  {Penny} A.~J.,  {Apps} K.,    {Carter} B.~D.,
  2002, \apj, 578, 565

\bibitem[\protect\citeauthoryear{{Chatterjee}, {Ford}, {Matsumura} \&
  {Rasio}}{{Chatterjee} et~al.}{2008}]{cha08}
{Chatterjee} S.,  {Ford} E.~B.,  {Matsumura} S.,    {Rasio} F.~A.,  2008, \apj,
  686, 580

\bibitem[\protect\citeauthoryear{{Croll}, {Matthews}, {Rowe}, {Gladman},
  {Miller-Ricci}, {Sasselov}, {Walker}, {Kuschnig}, {Lin}, {Guenther},
  {Moffat}, {Rucinski} \& {Weiss}}{{Croll} et~al.}{2007}]{cro07}
{Croll} B.,  {Matthews} J.~M.,  {Rowe} J.~F.,  {Gladman} B.,  {Miller-Ricci}
  E.,  {Sasselov} D.,  {Walker} G.~A.~H.,  {Kuschnig} R.,  {Lin} D.~N.~C.,
  {Guenther} D.~B.,  {Moffat} A.~F.~J.,  {Rucinski} S.~M.,    {Weiss} W.~W.,
  2007, \apj, 671, 2129

\bibitem[\protect\citeauthoryear{{Demory}, {Gillon}, {Barman}, {Bonfils},
  {Mayor}, {Mazeh}, {Queloz}, {Udry}, {Bouchy}, {Delfosse}, {Forveille},
  {Mallmann}, {Pepe} \& {Perrier}}{{Demory} et~al.}{2007}]{dem07}
{Demory} B.-O.,  {Gillon} M.,  {Barman} T.,  {Bonfils} X.,  {Mayor} M.,
  {Mazeh} T.,  {Queloz} D.,  {Udry} S.,  {Bouchy} F.,  {Delfosse} X.,
  {Forveille} T.,  {Mallmann} F.,  {Pepe} F.,    {Perrier} C.,  2007, \aap,
  475, 1125

\bibitem[\protect\citeauthoryear{{Dobbs-Dixon}, {Lin} \&
  {Mardling}}{{Dobbs-Dixon} et~al.}{2004}]{dob04}
{Dobbs-Dixon} I.,  {Lin} D.~N.~C.,    {Mardling} R.~A.,  2004, \apj, 610, 464

\bibitem[\protect\citeauthoryear{{Ford}, {Rasio} \& {Sills}}{{Ford}
  et~al.}{1999}]{for99}
{Ford} E.~B.,  {Rasio} F.~A.,    {Sills} A.,  1999, \apj, 514, 411

\bibitem[\protect\citeauthoryear{{Fressin}, {Guillot}, {Morello} \&
  {Pont}}{{Fressin} et~al.}{2007}]{fre07}
{Fressin} F.,  {Guillot} T.,  {Morello} V.,    {Pont} F.,  2007, \aap, 475, 729

\bibitem[\protect\citeauthoryear{{Gaudi}, {Seager} \& {Mallen-Ornelas}}{{Gaudi}
  et~al.}{2005}]{gau05}
{Gaudi} B.~S.,  {Seager} S.,    {Mallen-Ornelas} G.,  2005, \apj, 623, 472

\bibitem[\protect\citeauthoryear{{Guillot}, {Santos}, {Pont}, {Iro}, {Melo} \&
  {Ribas}}{{Guillot} et~al.}{2006}]{gui06}
{Guillot} T.,  {Santos} N.~C.,  {Pont} F.,  {Iro} N.,  {Melo} C.,    {Ribas}
  I.,  2006, \aap, 453, L21

\bibitem[\protect\citeauthoryear{{Halbwachs}, {Mayor} \& {Udry}}{{Halbwachs}
  et~al.}{2005}]{hal05}
{Halbwachs} J.~L.,  {Mayor} M.,    {Udry} S.,  2005, \aap, 431, 1129

\bibitem[\protect\citeauthoryear{{H{\'e}brard}, {Bouchy}, {Pont}, {Loeillet},
  {Rabus}, {Bonfils}, {Moutou}, {Boisse}, {Delfosse}, {Desort}, {Eggenberger},
  {Ehrenreich} \& {Forveille}}{{H{\'e}brard} et~al.}{2008}]{heb08}
{H{\'e}brard} G.,  {Bouchy} F.,  {Pont} F.,  {Loeillet} B.,  {Rabus} M.,
  {Bonfils} X.,  {Moutou} C.,  {Boisse} I.,  {Delfosse} X.,  {Desort} M.,
  {Eggenberger} A.,  {Ehrenreich} D.,    {Forveille} T.,  2008, \aap, 488, 763

\bibitem[\protect\citeauthoryear{{Hut}}{{Hut}}{1980}]{hut80}
{Hut} P.,  1980, \aap, 92, 167

\bibitem[\protect\citeauthoryear{{Hut}}{{Hut}}{1981}]{hut81}
{Hut} P.,  1981, \aap, 99, 126

\bibitem[\protect\citeauthoryear{{Jackson}, {Greenberg} \& {Barnes}}{{Jackson}
  et~al.}{2008}]{jac08}
{Jackson} B.,  {Greenberg} R.,    {Barnes} R.,  2008, \apj, 681, 1631

\bibitem[\protect\citeauthoryear{{Johnson}, {Marcy}, {Fischer}, {Laughlin},
  {Butler}, {Henry}, {Valenti}, {Ford}, {Vogt} \& {Wright}}{{Johnson}
  et~al.}{2006}]{joh06}
{Johnson} J.~A.,  {Marcy} G.~W.,  {Fischer} D.~A.,  {Laughlin} G.,  {Butler}
  R.~P.,  {Henry} G.~W.,  {Valenti} J.~A.,  {Ford} E.~B.,  {Vogt} S.~S.,
  {Wright} J.~T.,  2006, \apj, 647, 600

\bibitem[\protect\citeauthoryear{{Levrard}, {Winisdoerffer} \&
  {Chabrier}}{{Levrard} et~al.}{2009}]{lev09}
{Levrard} B.,  {Winisdoerffer} C.,    {Chabrier} G.,  2009, \apjl, 692, L9

\bibitem[\protect\citeauthoryear{{Mardling} \& {Lin}}{{Mardling} \&
  {Lin}}{2002}]{mar02}
{Mardling} R.~A.,  {Lin} D.~N.~C.,  2002, \apj, 573, 829

\bibitem[\protect\citeauthoryear{{Matsumura}, {Takeda} \& {Rasio}}{{Matsumura}
  et~al.}{2008}]{mat08}
{Matsumura} S.,  {Takeda} G.,    {Rasio} F.~A.,  2008, \apjl, 686, L29

\bibitem[\protect\citeauthoryear{{Mazeh}}{{Mazeh}}{2008}]{maz08}
{Mazeh} T.,  2008, in {Goupil} M.-J.,  {Zahn} J.-P.,  eds, EAS Publications
  Series Vol.~29 of EAS Publications Series, {Observational Evidence for Tidal
  Interaction in Close Binary Systems}.
pp 1--65

\bibitem[\protect\citeauthoryear{{Mazeh}, {Zucker} \& {Pont}}{{Mazeh}
  et~al.}{2005}]{maz05}
{Mazeh} T.,  {Zucker} S.,    {Pont} F.,  2005, \mnras, 356, 955

\bibitem[\protect\citeauthoryear{{Nagasawa}, {Ida} \& {Bessho}}{{Nagasawa}
  et~al.}{2008}]{nag08}
{Nagasawa} M.,  {Ida} S.,    {Bessho} T.,  2008, \apj, 678, 498

\bibitem[\protect\citeauthoryear{{Nordstr{\"o}m}, {Mayor}, {Andersen},
  {Holmberg}, {Pont}, {J{\o}rgensen}, {Olsen}, {Udry} \&
  {Mowlavi}}{{Nordstr{\"o}m} et~al.}{2004}]{nor04}
{Nordstr{\"o}m} B.,  {Mayor} M.,  {Andersen} J.,  {Holmberg} J.,  {Pont} F.,
  {J{\o}rgensen} B.~R.,  {Olsen} E.~H.,  {Udry} S.,    {Mowlavi} N.,  2004,
  \aap, 418, 989

\bibitem[\protect\citeauthoryear{{Pont}, {Melo}, {Bouchy}, {Udry}, {Queloz},
  {Mayor} \& {Santos}}{{Pont} et~al.}{2005}]{pon05}
{Pont} F.,  {Melo} C.~H.~F.,  {Bouchy} F.,  {Udry} S.,  {Queloz} D.,  {Mayor}
  M.,    {Santos} N.~C.,  2005, \aap, 433, L21

\bibitem[\protect\citeauthoryear{{Pont}, {Moutou}, {Bouchy}, {Behrend},
  {Mayor}, {Udry}, {Queloz}, {Santos} \& {Melo}}{{Pont} et~al.}{2006}]{pon06}
{Pont} F.,  {Moutou} C.,  {Bouchy} F.,  {Behrend} R.,  {Mayor} M.,  {Udry} S.,
  {Queloz} D.,  {Santos} N.,    {Melo} C.,  2006, \aap, 447, 1035

\bibitem[\protect\citeauthoryear{{Rasio} \& {Ford}}{{Rasio} \&
  {Ford}}{1996}]{ras96}
{Rasio} F.~A.,  {Ford} E.~B.,  1996, Science, 274, 954

\bibitem[\protect\citeauthoryear{{Ribas}, {Font-Ribera} \& {Beaulieu}}{{Ribas}
  et~al.}{2008}]{rib07}
{Ribas} I.,  {Font-Ribera} A.,    {Beaulieu} J.-P.,  2008, \apjl, 677, L59

\bibitem[\protect\citeauthoryear{{Saar}, {Butler} \& {Marcy}}{{Saar}
  et~al.}{1998}]{saa98}
{Saar} S.~H.,  {Butler} R.~P.,    {Marcy} G.~W.,  1998, \apjl, 498, L153+

\bibitem[\protect\citeauthoryear{{Strassmeier} \& {Hall}}{{Strassmeier} \&
  {Hall}}{1988}]{str98}
{Strassmeier} K.~G.,  {Hall} D.~S.,  1988, ApJS, 67, 453

\bibitem[\protect\citeauthoryear{{Udry}, {Mayor}, {Naef}, {Pepe}, {Queloz},
  {Santos} \& {Burnet}}{{Udry} et~al.}{2002}]{udr02}
{Udry} S.,  {Mayor} M.,  {Naef} D.,  {Pepe} F.,  {Queloz} D.,  {Santos} N.~C.,
    {Burnet} M.,  2002, \aap, 390, 267

\bibitem[\protect\citeauthoryear{{Winn}}{{Winn}}{2008}]{win08}
{Winn} J.~N.,  2008, ArXiv e-prints

\bibitem[\protect\citeauthoryear{{Winn}, {Johnson}, {Marcy}, {Butler}, {Vogt},
  {Henry}, {Roussanova}, {Holman}, {Enya}, {Narita}, {Suto} \& {Turner}}{{Winn}
  et~al.}{2006}]{win06}
{Winn} J.~N.,  {Johnson} J.~A.,  {Marcy} G.~W.,  {Butler} R.~P.,  {Vogt} S.~S.,
   {Henry} G.~W.,  {Roussanova} A.,  {Holman} M.~J.,  {Enya} K.,  {Narita} N.,
  {Suto} Y.,    {Turner} E.~L.,  2006, \apjl, 653, L69

\bibitem[\protect\citeauthoryear{{Zahn}}{{Zahn}}{1977}]{zah77}
{Zahn} J.-P.,  1977, \aap, 57, 383

\end{thebibliography}

\end{document}